\documentclass[usenatbib]{mn2e}
\usepackage{graphicx}

\begin{document}

\title[Superhumps: Theory and Observation]
{Superhumps: Confronting Theory with Observation}

\author[Pearson]
{K.\ J.\ Pearson
\\
Louisiana State University, Department of Physics and Astronomy,
Nicholson Hall, Baton Rouge, LA 70803-4001, USA. \\
}

\date{Accepted . Received ; 
in original form }

\maketitle

\begin{abstract} 
We review the theory and observations related to the ``superhump'' precession
of eccentric accretion discs in close binary sytems. We agree with
earlier work, although for different reasons, that the discrepancy between 
observation and dynamical theory implies that the effect of pressure in
the disc cannot be neglected. We extend earlier work that investigates this
effect to include the correct expression for the radius at which resonant
orbits occur. Using  analytic expressions for the accretion disc structure, we 
derive a relationship between the period excess and mass-ratio with the 
pressure effects included. This is compared to the observed data,
recently derived results for detailed integration of the disc equations and 
the equivalent empirically derived relations and used to predict values
for the mass ratio based on measured values of the period excess for 88 
systems.
\end{abstract}


\section{Introduction}
\label{sec:intro}

Cataclysmic Variables (CVs) are a class of close binary system, with a
typical orbital period of a few hours, where a 
white dwarf primary accretes material from its companion via Roche lobe 
overflow. 
The secondary is normally a late-type main sequence star although examples
with an evolved companion do exist (eg. GK~Per). In the absence of a 
significant white dwarf magnetic field, material arrives at
the primary after processing though an accretion disc. The dwarf nova 
subgroup show outbursts where the luminosity of the system increases by around
2--5~mag. Although not strictily periodic, these recur on a typical timescale
for each system ranging from tens of days to tens of years. The 
dwarf novae are further subdivided based on the properties of the
outbursts. We will be interested in the SU~UMa type where the systems show
occassional superoutbursts which have a brighter maximum ($\sim0.7$~mag.) and
longer duration ($\sim5$~times) than normal outbursts.

The most favoured explanation for dwarf nova outbursts involves an
ionization instability where the accretion disc mass increases until a
critical surface density 
\begin{equation}
\Sigma_{\rm max} = 114~\mbox{kg}~\mbox{m}^{-2} 
\left(\frac{r}{10^{8}~\mbox{m}}\right)^{1.05}  
M_{1}^{-0.35} \alpha_{C}^{0.86}
\end{equation}
is reached at some radius $r$. When this occurs, the disc switches 
into a ``hot'' state with higher viscosity
that causes a larger mass transport rate through the disc and increased 
luminosity. The disc
mass now steadily decreases until a second critical surface density 
\begin{equation}
\Sigma_{\rm min} = 82.5~\mbox{kg}~\mbox{m}^{-2} 
\left(\frac{r}{10^{8}~\mbox{m}}\right)^{1.05}  
M_{1}^{-0.35} \alpha_{H}^{0.8}
\end{equation}
is reached at some point. Upon meeting this condition, the disc transfers 
back to the ``cold'' quiescent state and the cycle repeats.
In these expressions $\alpha_{C}$ and $\alpha_{H}$ 
are the Shakura-Sunyaev viscosity 
parameters in the cold and hot states respectively and the 
primary mass $M_{1}$ is measured in solar masses \citep{cannizzo88}. 

SU~UMa superoutbursts also have the property of showing superhumps.
Here, an additional periodity ($P_{\rm sh}$) a few percent longer than the
orbital period ($P_{\rm orb}$) is apparent in the lightcurve. This 
is believed to arise from a precessing, eccentric accretion disc 
driven by a resonance between the orbiting disc material and the 
secondary. CVs in general and the SU~UMa systems in particular are 
well-reviewed by \citet{warner95}. Some systems other than dwarf novae also 
show superhumps which, by analogy, are believed to share a common origin with 
an eccentric disc now permanently present giving rise to 
``permanent superhumps'' \citep{retter00}. Similarly, some Low-Mass
X-ray Binaries (LMXBs); analogues to CVs where the primary is now a neutron 
star,
have also been discovered to have superhumps eg. KV~UMa \citep{zurita02}.
There is also a related phenomenon of ``negative'' superhumps which
occur on a period a few percent {\it shorter} than $P_{\rm orb}$. This is
believed to arise from precession of a disc warp and we will not consider these
in this paper.

The intent of this paper is to test our understanding of the 
(positive) superhump phenomenon by comparing observation to the predictions
of theoretical expressions for $P_{\rm sh}$. 
Since this is directly observable, if we can relate it to the funadmental parameters
of a system, we will have a method to indirectly measure such parameters. 

\section{Theoretical Background}
\label{sec:theback}

\citet{lubow91,lubow91b,lubow92} derived the final precession rate 
$\omega$ for an eccentric 
disc as the sum  of three terms: 
\begin{equation}
\omega=\omega_{\rm dyn} +\omega_{\rm press} + \omega_{\rm tran}
\end{equation}
where $\omega_{\rm dyn}$ is the dynamical precession frequency,
$\omega_{\rm press}$ is a pressure related term and $\omega_{\rm tran}$ is a 
transient term. This latter contribution is related to the time derivative
of the mode giving rise to the  dynamical precession. Thus, it 
may be important in the development phase of the superhumps but not
in steady state. As a result, we shall not consider
it in detail except to note that it can have either sign and thus can either
act to increase or decrease the precession rate. 

The dynamical term is the one arising from the resonance and is examined in
section~\ref{sec:dynthry}. The pressure term acts to slow the precession
and it is summarised in section~\ref{sec:presthry}.

\subsection{Dynamical Precession Theory}
\label{sec:dynthry}

\citet{hirose90} derived the general expression for the ratio of the dynamical
disc precession $\omega_{\rm dyn}$ and orbital $\omega_{\rm orb}$ frequencies 
in terms of the mass ratio and radius of disc material.
Their equation (8) is
\begin{equation}
\frac{\omega_{\rm dyn}}{\omega_{\rm orb}} =
\frac{q}{\left(1+q\right)^{\frac{1}{2}}}
\left[\frac{1}{2r^{\frac{1}{2}}} \frac{d}{d\!r}
\left(r^{2} \frac{d\!B_{0}}{d\!r}\right)\right]
\label{eqn:osaki8}
\end{equation}
where
\begin{equation}
B_{0}(r) = \frac{1}{2}b^{0}_{\frac{1}{2}} = F\left(\frac{1}{2},\frac{1}{2},1,
r^{2}\right)
\end{equation}
\citep{brumberg95} is the zeroth order Laplace coefficient given in terms of 
the hypergeometric function $F$, $q=M_{2}/M_{1}~(<1)$ is the mass-ratio
and $r$ is the radius of 
orbiting material expressed as a fraction of the separation $d$. 
This evaluates to
\begin{equation}
\frac{\omega_{\rm dyn}}{\omega_{\rm orb}} =
\frac{3}{4}\frac{q}{\left(1+q\right)^{\frac{1}{2}}}
r^{\frac{3}{2}}\sum_{n=1}^{\infty} a_{n} r^{2(n-1)}
\label{eqn:omrat}
\end{equation}
where the coefficients are given by
\begin{equation}
a_{n}=\frac{2}{3}(2n)(2n+1)\prod_{m=1}^{n} \left(\frac{2m-1}{2m}\right)^{2}
\label{eqn:coeffrel}
\end{equation}
\citep{pearson03}.
\citet{lubow92} used the fixed value of $r=0.477$ enigmatically described
as ``corrected for the presence of the companion'' and thus presumably
in the limit of $q\rightarrow0$. 
\citet{FKR}, however, give the radius for
 $j$:$j-1$ resonances as,
\begin{equation}
r_{j}
=\frac{1}{j^{\frac{2}{3}}\left(1+q\right)^{\frac{1}{3}}}.
\label{eqn:resrad}
\end{equation}
This evaluates to r=0.481 for the case of $j=3$ and vanishing $q$; very 
close to
the value of the other paper but retaining accuracy for $q\neq0$.  
Substituting into (\ref{eqn:omrat}) gives
\begin{equation}
\frac{\omega_{\rm dyn}}{\omega_{\rm orb}}  = 
\frac{3}{4j}\frac{q}{1+q} 
\sum_{n=1}^{\infty} 
\frac{a_{n}}{\left[j^{2}{(1+q)}\right]^{\frac{2(n-1)}{3}}}.
\label{eqn:omratfull}
\end{equation}
The canonical approximation
\begin{equation}
P_{\rm dyn}\approx\frac{3.85(1+q)}{q}P_{\rm orb}
\end{equation}
\citep{warner95},  
is recovered by setting $j=3$ and evaluating the summation with $q=0.16$. 
The limiting mass 
ratio $q\approx0.22$ found by \citet{whitehurst88a} arises from the largest 
value for which $r_{3}$ remains within the last stable stream line
\citep{molnar92}. Numerical simulations, however, still produce identifiable 
superhumps up to a mass ratio of $q\approx0.33$ \citep{whitehurst94,murray00b}.
\citet{whitehurst91} used an approximation for the disc tidal radius
\begin{equation}
R_{\rm T}\approx\beta R_{\rm L,1}
\label{eqn:tidalrad}
\end{equation}
with $\beta\approx0.9$. When coupled to Eggleton's 
formula \citep{eggleton83} for the primary's Roche lobe radius 
\begin{eqnarray}
R_{\rm L,1} & = &\frac{0.49q^{-\frac{2}{3}}}{0.6q^{-\frac{2}{3}}
+\ln(1+q^{-\frac{1}{3}})}\\
 & \equiv & E(q^{-1}) \label{eqn:eggformula}
\end{eqnarray}
and equated to the 3:2 resonance radius in equation~(\ref{eqn:resrad}), this
gives a limiting mass ratio of $q_{\rm max}=0.28$, although this is sensitive 
to the 
choice of $\beta$. It should be noted, however, that this differs from the 
often cited value of $q_{\rm max}=0.33$ quoted in that paper, as equation (5) 
there contains an incorrect power of $q$. A slightly less {\it ad hoc}
expression for the tidal radius comes from fitting to the simulations
of \citet{paczynski77}
\begin{equation}
R_{\rm T}=\frac{0.60}{1+q}~~~~~~~~~~~0.03<q<1.
\end{equation}
Equating this to equation~(\ref{eqn:resrad}) gives a 
limiting mass ratio of
\begin{equation}
q=(0.6)^{\frac{3}{2}}j-1
\label{eqn:qmax}
\end{equation}
which sets a maximum $q_{\rm max}=0.39$ for a 3:2 resonance.

\subsection{Pressure Contribution}
\label{sec:presthry}

\citet{lubow92} showed that the pressure term can be expressed as 
\begin{equation}
\omega_{\rm press}=-\frac{k^{2} c^{2}}{2\omega_{\rm p}}
\label{eqn:ompress1}
\end{equation}
where $\omega_{\rm p}$ is the angular orbital frequency of a parcel of gas
in the disc,
$k$ is the radial wavenumber of the mode and $c$ is the gas sound speed.
Clearly the pressure term acts in the opposite, retrograde sense to the
dynamical term.
For a spiral wave, the pitch angle $i$ is related to $k$ by
\begin{equation} 
\tan  i =\frac{1}{k r}.
\label{eqn:kirel}
\end{equation}
From the resonance condition \citep[eg.][eqn. 3.37]{warner95}, we have
\begin{equation}
(j-1)(\omega_{\rm p}-\omega)=j(\omega_{\rm p}-\omega_{\rm orb})
\end{equation}
which becomes, when $\omega\ll\omega_{\rm orb}$,
\begin{equation}
\omega_{p}=j\omega_{\rm orb}.
\label{eqn:ompapprox}
\end{equation}
Hence, for the 3:2 resonance $\omega_{\rm p}=3\omega_{\rm orb}$. For the fixed 
radius
$r=0.477$, \citet{montgomery01} corrected earlier errors to derive 
a contribution 
\begin{equation}
\omega_{\rm press}=-0.7325\omega_{\rm orb}
\left(\frac{c}{\omega_{\rm orb} d} \frac{1}{\tan i}\right)^{2}.
\end{equation}
For our general case using (\ref{eqn:ompapprox}) and  where $r$ is 
given by equation~(\ref{eqn:resrad}), we have
\begin{equation}
\omega_{\rm press}=-\frac{j^{\frac{1}{3}}}{2}(1+q)^{\frac{2}{3}}
\omega_{\rm orb} \left(\frac{c}{\omega_{\rm orb} d} \frac{1}{\tan i}
\right)^{2}.
\label{eqn:ompress}
\end{equation}
To proceed further, we need to understand the behaviour of the final 
dimensionless term in brackets. Since values for  $c$ and $i$
may vary according to the peculiar characteristics of any particular
system, we will examine this in more detail in section~\ref{sec:presscomp}.

\section{Comparison}
\label{sec:comp}

Observers normally present their measurements of the precesion period
in terms of the period excess 
\begin{equation}
\epsilon = \frac{P_{\rm sh}-P_{\rm orb}}{P_{\rm orb}} 
\end{equation}
or equivalently (noting $\omega_{\rm sh}=\omega_{\rm orb}-\omega$)         
\begin{eqnarray}
\epsilon & = & \frac{\omega}{\omega_{\rm orb}-\omega} \\
         & = & 
\left[\left(\frac{\omega_{\rm dyn}+\omega_{\rm press}}{\omega_{\rm orb}}
\right)^{-1}-1\right]^{-1}\\
         &\approx&
\frac{\omega_{\rm dyn}+\omega_{\rm press}}{\omega_{\rm orb}}
~~~~~~~~~~~~~~\mbox{if~}\omega_{\rm dyn},\omega_{\rm press}\ll\omega_{\rm orb}.
\end{eqnarray}

Since there are relatively few systems with accurately measured values
of $q$, it is often convenient to make use of the
theoretical relation 
\begin{equation}
M_{2}\approx0.11P_{\rm orb}
\end{equation}
\citep{FKR}
that follows from the assumption that the secondary has a main sequence
structure or an observationally derived equivalent
\begin{equation}
M_{2}=(0.038\pm0.003)P_{\rm orb}^{1.58\pm0.09}
\label{eqn:m2prel}
\end{equation}
\citep{smith98}. In both cases $M$ is measured in solar masses and 
$P_{\rm orb}$ in hours. 

\subsection{Dynamical precession only}
\label{sec:dynonly}

\begin{figure}
\begin{center}

\includegraphics [angle=270,scale=0.5]{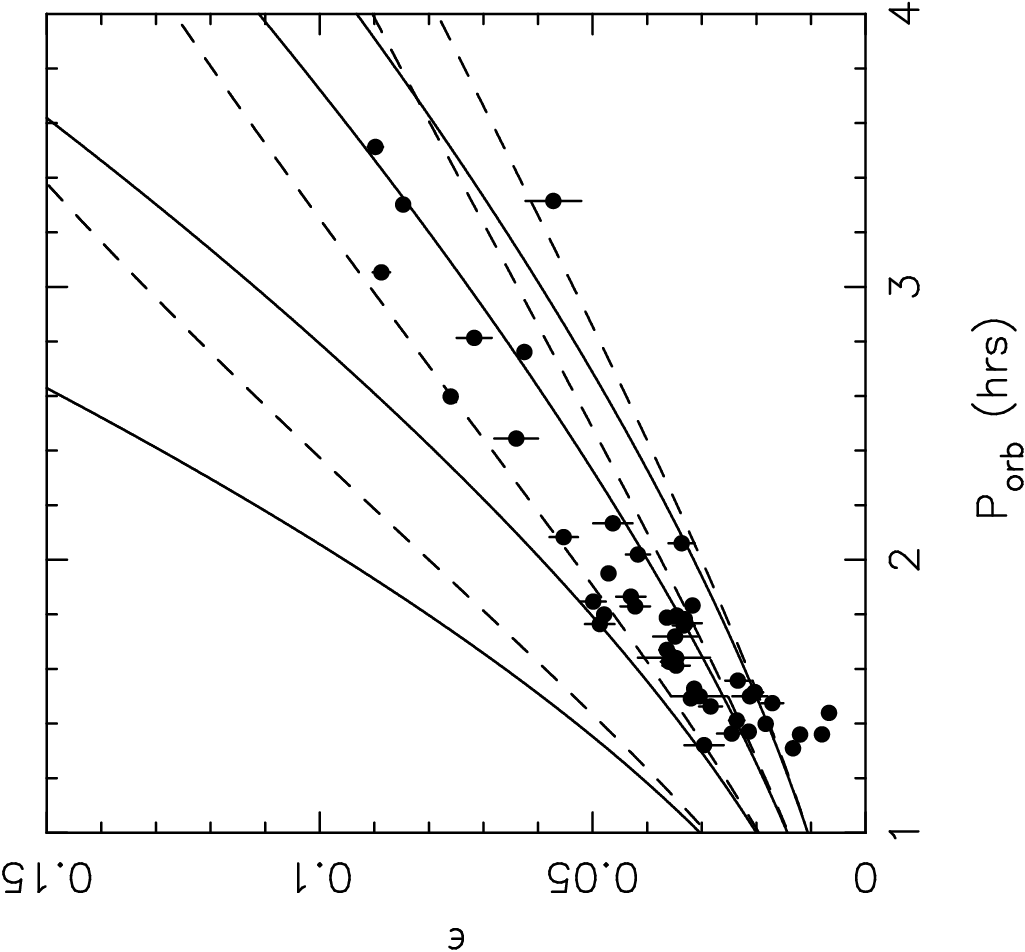}
\caption{Comparison of the observed superhump data from 
\protect\citet{patterson98} with the model from \citet{murray00}, where the
resonant radius is independent of $q$ (solid), and with a model including 
the $q$ dependance (dashed). Both family of curves show the theoretical
lines derived using the mean $(0.76)$ and $\pm1\sigma$ $(\pm0.22)$ values for 
$M_{1}$ of all CVs and in the 
$M_{2}$--$P_{\rm orb}$ relation (equation \protect\ref{eqn:m2prel}) 
from \protect\citet{smith98} as well as a further line with $M_{1}=1.44$. 
$\epsilon$ decreases with increasing $M_{1}$. The turn to lower $\epsilon$
at $P_{\rm orb}< 1.4$ may arise from the secondary becoming degenerate
and deviating from a main sequence structure as assumed for the theoretical 
curves.}

\protect\label{fig:epsporb}
\end{center}

\end{figure}

For different assumed values of $M_{1}$ we can 
plot theoretically predicted lines on the $\epsilon$--$P_{\rm orb}$ plane. 
\citet{murray00} used equation~\ref{eqn:omrat} with the fixed radius
value given by \citet{lubow92} to compare the observed distribution with 
theory in just this way. He concluded that 
``superhumps observations {\it cannot} be adequately explained in terms of
purely dynamical precession''. However, including the $q$ dependence of $r$ 
given in equation~\ref{eqn:omratfull} leads us to a different conclusion. 
Figure~\ref{fig:epsporb} reproduces the comparison of \citet{murray00} using 
both methods. The lines are plotted for $M_{1}=0.76$, $0.76\pm0.22$ and 
$M_{1}=1.44$. This shows that the distribution is compatible with
the boundary imposed by the condition $M_{\rm wd}<1.44$
when the full $q$ dependence is included. In fact, the most we can conclude is
that the distribution of superhumping systems suggests that they have a primary
mass higher than the general CV population. This would not be entirely
suprising since to be a superhumping system requires a small mass ratio and
will thus tend to select higher $M_{1}$ systems. Also apparent is the deviation
of the secondary from a main sequence structure at the short period turnoff.

\begin{figure}
\begin{center}
\includegraphics [angle=270,scale=0.5] {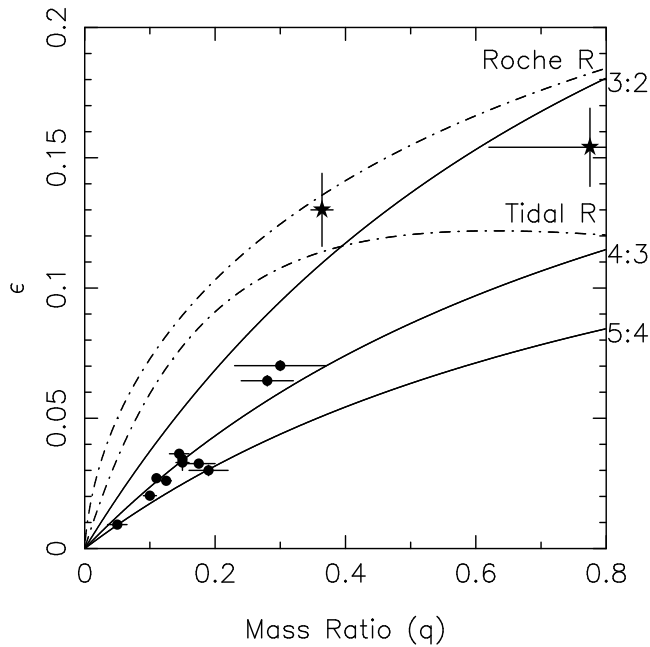}
\caption{Comparison of the observed superhump data from 
\protect\citet{patterson05} (circles), for systems with accurately  
measured values of $q$, and the dynamical precession rates of discs with radii
calculated under different assumptions. The resonance period excesses are 
calculated using equation~(\ref{eqn:omratfull}) and the discs with radii
equal to the tidal and (unphysically) the Roche lobe radius are calculated 
using equation~(\ref{eqn:omrat}). The stars mark the additional `challenging' 
systems U~Gem and TV~Col.
}
\protect\label{fig:epsofq}
\end{center}

\end{figure}

\begin{figure}
\begin{center}
\includegraphics [angle=270,scale=0.5] {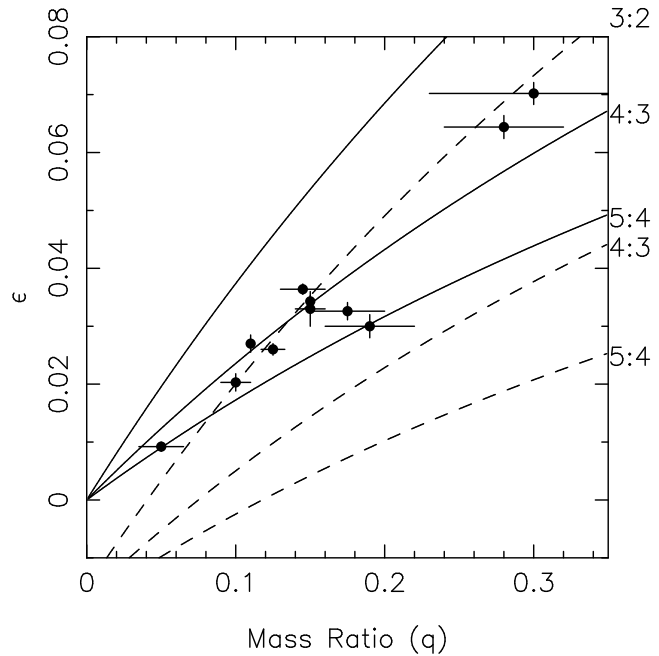}
\caption{A closer view of the data used  in the comparison of the observed 
superhump data from 
\protect\citet{patterson05}, for systems with accurately  
measured values of $q$, and the dynamical precession rates of discs
at resonant radii (solid line). Also plotted are the model lines including an
approximation to pressure effect (Model A; dashed line).
}
\protect\label{fig:epsofqzoom}
\end{center}

\end{figure}

More convincing evidence for the need for a pressure related effect on
the precession rate is provided by the data for $\epsilon$ and $q$ 
plotted in Figure~\ref{fig:epsofq}. The 11 systems on this plot are those
used in \citet{patterson05}. These have directly determined values of $q$.
The exception is OY~Car which appears in Fig.~9 of that paper but not
in the corresponding Table~7. We take the values for this system from 
\citet{patterson01}. We can see
that the data are certainly not compatible with the asumption of a 3:2
resonance. If anything, they cluster around the prediction for a 4:3
resonance, although given the way the resonances `pile up' one can argue that 
it is inevitable that some resonance would fall near the data. It is 
intriguing to note however,
just how close the two most accurately measure values of $q$ for the systems
XZ~Eri and DV~UMa \citep{feline04b} lie to the 4:3 theoretical line 
(see Figure~\ref{fig:epsofqzoom}). 
It would thus be extremely interesting to see modelling using the 
method employed by 
these authors applied to the other eclipsing systems to determine similarly
accurate values for $q$. We summarise the
$\chi^{2}$ value for the data against each resonance in 
Table~\ref{tab:reschi2} and note the remarkable reduction in $\chi^{2}$ for
the 4:3 model. If a purely dynamical 4:3 resonance were the correct model,
the probability that the value of $\chi^{2}$ would exceed the measured value 
is 0.23.

\begin{table}
\begin{center}
\begin{tabular}{cc}
\hline
Resonance & $\frac{\chi^{2}}{N_{\rm obs}}$ \\ \hline
3:2 &  49.5 \\
4:3 &  1.28 \\
5:4 &  18.9 \\ \hline
\end{tabular}
\caption{Summary of the $\chi^{2}$ statistic comparing purely dynamical 
resonant precession to the observed period excess.}
\end{center}
\protect\label{tab:reschi2}
\end{table}

We have followed the precendent of \citet{patterson05} in the systems 
considered above but we mention here briefly two further systems: U~Gem and
TV~Col. On the grounds that U~Gem does not show superhumps, \citet{patterson05}
used its observed mass ratio to place an upper limit on the 
system BB~Dor which does have measured period excess. However, \citet{smak04}
reported the detection of superhumps in the 1984 outburst of U~Gem. As a 
result, we dropped BB~Dor from the analysis. Similarly, \citet{retter03}
report a superhump detection near the expected 6.3-h in TV~Col from an 
exhaustive search in archival and from fresh 2001 observations. Two other 
campaigns, however, failed to confirm this result \citep{patterson05}. 
Neither U~Gem nor TV~Col fit comfortably within the context
of the precessing disc theory set out above. However, both have
noticeably large errors on the value of their period excess. In the case
of U~Gem this reflects a systematic trend as the putative superhump period
drifted to longer values over time. The outburst was also notable for its 
unusual length and completely out of character for the system. To reach
the observed period excess would require the disc to extend beyond the 
tidally truncated radius, which is not impossible for a disc of finite
viscosity \citep{lyndenbell74,papaloizou77}, but is uncomfortably close to the 
value we would get if the 
disc had a radius equal to that of the primary's entire Roche lobe! The most
plausible explanation for such a large period excess and the unusual 
character is that the transient
term $\omega_{\rm tran}$ has become important and that the change in the 
observed period excess reflects a change in the rate of growth of the disc 
eccentricity. The uncertainties in the values for TV~Col make it a difficult
system to assess. The expected disc radius is again much larger than the tidal
truncation radius and in addition to the large range of allowed $q$ there is 
the possibility of unknown systematic errors due to the multiple components
of the emission lines \citep{hellier93}. We thus neglect this system also
but echo the call of \citet{retter03} for a search for permanent superhumps in
similarly long period systems.

Despite the encouraging agreement with a 4:3 resonance there are several 
theoretical hurdles to overcome before we could accept such an explanation. 
The theory outlined in Section~\ref{sec:theback} is only a summary of the 
extensive literature in this area
(eg \citet{goldreich78,goldreich79,borderies83}). 
In outline, the analytical method employed is to decompose the effective 
potential into its 
harmonic components, to introduce this into the fluid equations in a 
standard linear perturbation analysis and to look at the response of the disc 
material. The behaviour of the disc in superhumping
systems has a close analogue in galactic dynamics. The ``dynamical resonance''
above corresponds to the inner Lindblad resonance of the system and there is
a similar result to that given, for example by \citet{binneybook}, that 
leads to spiral waves being generated in the disc. 
It is these spiral waves that couple with the tides to excite disc 
eccentricity. There is also
a corotational resonance that acts to suppress eccentricity.

Cursory consideration shows the fundamental contradiction of treating the 
dissipation arising from an eccentric disc (an inherently collective 
phenomenon) by the precession of single particle orbits at a resonant 
radius as carried out above. As a first attempt to correct for the untreated 
collective effects, we might assume that the precession can be 
characterised by an effective radius ($r_{\rm eff}$) interior to $r_{3}$ that 
would produce dynamical precession with the observed period. 
The values for $r_{\rm eff}$ derived from the observed superhump period
excesses of the \citet{patterson05} systems with well measured $q$ are
given in Table~\ref{tab:reff}. Assuming that the ratio $r_{\rm eff}/r_{3}$
is a constant for all systems, we derive a best value for it of 0.827. 
This places $r_{\rm eff}$ extremely close to $r_{4}$ with a barely different 
value for the total $\chi^{2}$. In fact, the radius calculated from a $j=4$ 
resonance differs from this best possible radius by a suprisingly small 0.2\%!

\begin{table}
\protect\label{tab:reff}
\begin{center}
\begin{tabular}{lcccccc} \hline
System &  $\frac{r_{\rm eff}}{r_{3}}$ \\\hline
 WZ Sge &  0.718  \\
 OY Car &  0.770  \\
 XZ Eri &  0.844  \\
 IY UMa &  0.787  \\
 Z  Cha &  0.864  \\
 HT Cas &  0.816  \\
 DV UMa &  0.831  \\
 OU Vir &  0.762  \\
 V2051 Oph & 0.707\\
 DW UMa &  0.873  \\
 UU Aqr &  0.886  \\ \hline
\end{tabular}
\caption{Values for $r_{\rm eff}$ derived from the observed period excesses 
for the \protect\citet{patterson05} calibration systems.} 
\end{center}
\end{table}

The above result not withstanding, a treatment that deals explictly with
the coupling of the tides, spiral arms and disc eccentricity explicitly
ought to be preferred. This is exactly the approach used by \citet{lubow91}
to derive the additional term related to pressure effects which we turn to 
next. One of the important results from that paper was the recognition that
the 3:2 resonance ``is unique in that it is the innermost resonance
for which an eccentric Lindblad resonance appears without an overlapping
eccentric corotational resonance\ldots This property allows that resonance to 
easily excite eccentricity.'' Such a conclusion is a strong argument
against any dynamical resonance other than 3:2 being an important factor in 
the disc behaviour. The ability of a secondary
magnetic field to give rise to a precession rate characteristic of a 
higher resonance \citep{pearson97,pearson03} may reflect the fact that the 
perturbation in that case 
is no longer small and capable of represention by a linear analysis.

\subsection{Inclusion of Pressure Effects}
\label{sec:presscomp}

Proceeding under the assumption that the excited resonance is 3:2 and that 
the difference between the measured and expected $\epsilon$ is due to the
pressure effect we can update the analysis of \citet{murray00} for all the
systems with measured $q$. These are summarised in table~\ref{tab:press}.
For the final column we have used values for $M_{1}$ from \citet{patterson05}
except for V2051 Oph which we take from \citet{rkcat}.

\begin{table*}
\protect\label{tab:press}
\begin{center}
\begin{tabular}{lcccccc} \hline
System & $\omega$ & $\omega_{\rm dyn}$ &
$\omega_{\rm press}$ & $\sqrt{2\eta_{\rm A}}$ 
& $\frac{c}{\omega_{\rm orb} d}$ & $c$ \\ 
       & (d$^{-1}$) & (d$^{-1}$) & (d$^{-1}$) &  
& & ($10^{4}$ m s$^{-1}$) \\\hline
 WZ Sge &    1.01  &  2.13 &   -1.12 &    0.12 &   0.036 & 2.07 \\
 OY Car &    1.98  &  3.59 &   -1.61 &    0.14 &   0.044 & 2.16 \\
 XZ Eri &    2.70  &  4.02 &   -1.32 &    0.13 &   0.039 & 2.02 \\
 IY UMa &    2.15  &  3.71 &   -1.56 &    0.15 &   0.047 & 2.29 \\
 Z  Cha &    2.96  &  4.17 &   -1.21 &    0.13 &   0.041 & 1.80 \\
 HT Cas &    2.73  &  4.34 &   -1.62 &    0.15 &   0.047 & 2.14 \\ 
 DV UMa &    2.43  &  3.73 &   -1.30 &    0.15 &   0.046 & 2.38 \\
 OU Vir &    2.73  &  4.99 &   -2.26 &    0.18 &   0.055 & 2.87 \\
 V2051 Oph & 2.93  &  6.20 &   -3.27 &    0.20 &   0.061 & 3.21 \\
 DW UMa &    2.78  &  3.79 &   -1.01 &    0.16 &   0.049 & 1.99 \\
 UU Aqr &    2.52  &  3.33 &   -0.81 &    0.16 &   0.048 & 1.78 \\ \hline
\end{tabular}
\caption{Derived pressure force contribution to the precession of the
\protect\citet{patterson05} calibration systems. The last two columns 
assume $i=17^{\circ}$.} 
\end{center}
\end{table*}

To make progress with our comparison we need values for $c$ and $i$ in equation
\ref{eqn:ompress}. We will consider two cases: A) where 
$\frac{c \cot i}{\omega_{\rm orb} d}$  is fixed for all systems and  
B) where $c$ is evaluated using  analytic expression taken from detailed
disc models.

\subsubsection{Model A}

If we assume that the final bracketed term of equation~(\ref{eqn:ompress})
is constant we can rewrite it as
\begin{equation}
\frac{\omega_{\rm press}}{\omega_{\rm orb}}=-j^{\frac{1}{3}} \eta_{\rm A} 
\left(1+q\right)^{\frac{2}{3}}
\end{equation}
where
\begin{equation}
\eta_{\rm A}=\frac{1}{2}\left(\frac{c \cot i}{\omega_{\rm orb} d} \right)^{2}.
\end{equation}
\citet{lubow92} gave a range of 0.01--0.05 for the ratio 
$\frac{c}{\omega_{\rm orb} d}$. \citet{murray00} used a value of 0.05
in his considerations to derive values for $i$. Using this value and the mean 
$i=17^\circ$ found from simulations by \citet{montgomery01}, we have 
$\eta_{\rm A}=0.0134$. Compared to the data for the 
eclipsing systems, this 
gives $\frac{\chi^{2}}{N_{\rm obs}}=3.3$. Allowing $\eta_{\rm A}$
to be a free parameter, however, we derive an optimal value of 
$\eta_{\rm A}=0.0107$ with a corresponding
$\frac{\chi^{2}}{N_{\rm obs}-1}=1.42$. Figure \ref{fig:epsofqzoom} shows
the comparison for this value in graphical form.

\begin{figure}
\begin{center}
\includegraphics [angle=270,scale=0.5] {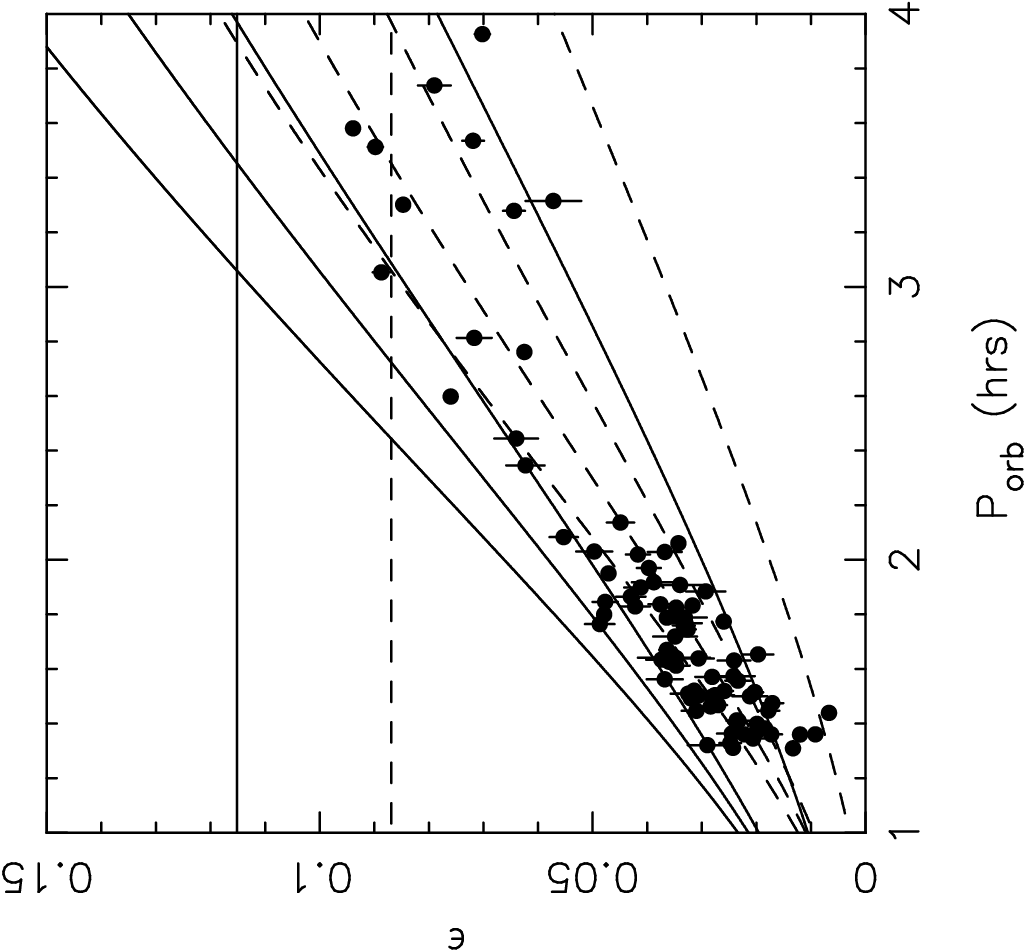}
\caption{
Comparison of the observed superhump data from 
\protect\citet{patterson05} with a model with purely dynamical precession 
(solid) and including the pressure effect (dashed). Both family of curves 
show the theoretical
lines derived using the mean ($0.69$) and $\pm1\sigma$ ($0.01$) values for 
$M_{1}$ found for {\it dwarf novae} and in the 
$M_{2}$--$P_{\rm orb}$ relation (equation \protect\ref{eqn:m2prel}) 
from \protect\citet{smith98} as well as a further line with $M_{1}=1.44$. 
$\epsilon$ decreases with increasing $M_{1}$. The horizontal lines mark the
limiting precession rate derived from $q_{\rm max}=0.39$ in either 
case. 
}
\protect\label{fig:epspcomp}
\end{center}
\end{figure}

The corresponding plot to Figure~\ref{fig:epsporb}, comparing a purely 
dynamical 3:2 resonance and a model including pressure, is shown in 
Figure~\ref{fig:epspcomp}. It should be noted here that the values assumed
for $M_{1}$ in producing the lines come from the weighted mean derived 
by \citet{smith98} for {\it dwarf nova} systems ($M_{1}=0.69\pm0.01$) rather 
than that for all systems used by \citet{murray00}. We have also updated the
data to the latest compilation from \citet{patterson05}. Coupling the mean 
primary mass
with $q_{\rm max}=0.39$, we can derive an equivalent $P_{\rm orb,max}=3.5$-h.
For the ultimate restriction $M_{1}<1.44M_{\odot}$ we have
$P_{\rm orb,max}=5.5$-h (coincidentally the same as that of TV~Col).

We might like to compare the implied distribution of $M_{1}$ from our
model with that of \citet{smith98}. However, this is
precluded by the selection effect alluded to in 
section~\ref{sec:dynonly} being at work.
Systems with longer periods (and thus higher $M_{2}$) can accomodate more 
massive primaries and still fit within the $q_{\rm max}$ limitation. Hence,
the distribution of $M_{1}$ would be expected to differ between all dwarf
novae and the superhumping subset. Figure~\ref{fig:epspcomp}
also shows systems lieing above the limiting $\epsilon$ allowed by our 
formulation of the pressure term. This, along with the range of derived 
$\eta_{\rm A}$ shown in Table~\ref{tab:press} and the poorer fit to the data than
a simple resonance model, suggests that a one size fits all
value for $\eta$ is not appropriate.

\subsubsection{Model B}

Since the effect of the pressure term relies on the sound speed in the disc,
we turn to detailed models of hot discs to evaluate this in terms of
fundamental parameters. From equation~A1 of \citet{cannizzo92a} we have
the mid-plane temperature in terms of the mass transport rate through the disc
\begin{equation}
T_{\rm mid}=\left(\frac{64}{9}\frac{\sigma}{\kappa_{0}}\right)^{-\frac{1}{10}}
\left(\frac{\mu_{\rm H}}{R}\right)^{\frac{1}{4}} 
\omega_{\rm p}^{\frac{1}{2}} 
\alpha_{\rm H}^{-\frac{1}{5}} \left(\frac{\dot{M}}{2\pi}\right)^{\frac{3}{10}}
\label{eqn:tmid}
\end{equation}
where $\sigma$ is the Stefan-Boltzman constant, $R$ is the gas constant, 
$\alpha_{\rm H}$ and $\mu_{\rm H}$ are the Sunyaev-Shakura viscosity parameter 
and mean molecular weight in the hot state respectively and assuming
an opacity $\kappa=\kappa_{0} \rho T^{-3.5}$ where 
$\kappa_{0}=2.8\times10^{23}~\mbox{m}^{2}~\mbox{kg}^{-1}$ is appropriate
\citep{cannizzo88}.
We can find a suitable value of $\dot{M}$ for an outbursting dwarf
nova using the approach of \citet{cannizzo93} and \citet{cannizzo88} by 
assuming that the 
disc fills to a mass $fM_{\rm max}$. $M_{\rm max}$ is the  maximum mass
the disc can hold in the  cold state without exceeding $\Sigma_{\rm max}$ at
some point ie.
\begin{eqnarray}
fM_{\rm max} & = & f \int_0^{r_{\rm  d}} 2 \pi r \Sigma_{\rm max}(r) d\,r 
\\
           & = & f \frac{2\pi r_{\rm d}^{2}}{3.05} \Sigma_{\rm max}(r_{\rm d}).
\label{eqn:fmmax}
\end{eqnarray}
Now, equation~A3 of \citet{cannizzo92a} gives the surface density in the  hot
state as
\begin{eqnarray}
\Sigma & = & \left(\frac{64}{9}\frac{\sigma}{\kappa_{0}}\right)^{\frac{1}{10}}
\left(\frac{\mu_{\rm H}}{R}\right)^{\frac{3}{4}}
\omega_{\rm p}^{\frac{1}{2}} 
\alpha_{\rm H}^{\frac{4}{5}} \left(\frac{\dot{M}}{2\pi}\right)^{\frac{7}{10}}
\\ \nonumber
       & = & 405~\mbox{kg}~\mbox{m}^{-2} 
\mu_{\rm H}^{\frac{3}{4}}
\alpha_{\rm H}^{-\frac{4}{5}} 
M_{1}^{\frac{1}{4}} 
\left(\frac{r}{10^{8}~\mbox{m}}\right)^{-\frac{3}{4}} \\ 
& & \times 
\left(\frac{\dot{M}}{10^{-10}~\mbox{M}_{\odot}~{\mbox{y}^{-1}}}
\right)^{\frac{7}{10}}
\end{eqnarray}
where we have used 
$\omega_{\rm p}=\left(GM_{1}M_{\odot}r^{-3}\right)^{\frac{1}{2}}$.
Integrating this and equating it to the expression for $fM_{\rm max}$ from
(\ref{eqn:fmmax}), we can rearrange for the mass transport rate through the
disc
\begin{eqnarray}
\dot{M} & = & 9.67\times10^{-8}~\mbox{kg}~\mbox{s}^{-1}~ 
\left(\frac{\alpha_{\rm H}}{0.1}\right)^{1.14}
\left(\frac{\alpha_{\rm C}}{0.02}\right)^{-1.23}
\mu_{\rm H}^{-1.07} \nonumber \\  & & \times
r_{\rm d}^{2.57}
M_{1}^{-0.86}
\left(\frac{f}{0.4}\right)^{1.43}
\end{eqnarray}
which, with equation~(\ref{eqn:tmid}), gives
\begin{eqnarray}
c^{2} & = & \frac{\gamma k T}{\mu_{H} m_{\rm H}} \\
 & = & 5.66\times10^{4}~\mbox{m}^{1.229}~\mbox{s}^{-\frac{3}{2}}
\left(\frac{\alpha_{\rm H}}{0.1}\right)^{0.142}
\left(\frac{\alpha_{\rm C}}{0.02}\right)^{-0.369} \nonumber \\ 
&& \times \mu_{\rm H}^{-1.071}
r_{\rm d}^{0.771}
M_{1}^{-0.258}
\left(\frac{f}{0.4}\right)^{0.429}
\omega_{p}^{\frac{1}{2}}.
\end{eqnarray}
Finally, we can combine this with equations~(\ref{eqn:resrad}),
(\ref{eqn:ompress1}), (\ref{eqn:kirel}) and (\ref{eqn:ompapprox}) to get
\begin{eqnarray}
\frac{\omega_{\rm press}}{\omega_{\rm orb}} & = &
-2.83\times10^{4} j^{\frac{5}{6}} (1+q)^{\frac{2}{3}} \cot^{2} i
\left(\frac{\alpha_{\rm H}}{0.1}\right)^{0.142} \nonumber \\  
& & \times
\left(\frac{\alpha_{\rm C}}{0.02}\right)^{-0.369}
\mu_{\rm H}^{-1.071}
r_{\rm d}^{0.771}
M_{1}^{-0.258} \nonumber \\  & & \times
\left(\frac{f}{0.4}\right)^{0.429}
\omega_{\rm orb}^{-\frac{3}{2}}.
\label{eqn:finala}
\end{eqnarray}
Since we want an expression for $\omega_{\rm press}$ in terms of $q$ only,
we look to eliminate $\omega_{\rm orb}$ by using a form of the mass-radius 
relation recommended by \citet{smith98}
\begin{equation}
\frac{R_{2}}{R_{\odot}}= (0.91\pm0.09)M_{2}^{0.75\pm0.04}.
\label{eqn:massradrel}
\end{equation}
When this is equated to the size of the secondary's Roche lobe from 
(\ref{eqn:eggformula}), we can rearrange for the separation
\begin{equation}
d=\frac{0.91M_{2}^{0.75} R_{\odot}}{E(q)}.
\end{equation}
Using this with Kepler's Law
\begin{equation} 
\omega_{\rm  orb}=\left[GM_{1}(1+q)M_{\odot}\right]^\frac{1}{2} 
d^{-\frac{3}{2}}
\end{equation}
and a disc radius $r_{\rm d}=\beta R_{\rm L,1}$, we arrive at the final
expression
\begin{eqnarray}
\frac{\omega_{\rm press}}{\omega_{\rm orb}} & = & -j^{\frac{5}{6}}
\eta_{\rm B} \frac{\left[E(q^{-1})\right]^{0.771}}
{\left[E(q)\right]^{1.021}}
\frac{q^{0.766}}{(1+q)^{\frac{1}{12}}} \label{eqn:finompress}
\end{eqnarray} 
where
\begin{eqnarray}
\eta_{\rm B} & = & 0.0209 \cot^{2} i
\left(\frac{\alpha_{\rm H}}{0.1}\right)^{0.142}
\left(\frac{\alpha_{\rm C}}{0.02}\right)^{-0.369} \nonumber \\
& &
\times \mu_{\rm H}^{-1.071} 
 \left(\frac{f}{0.4}\right)^{0.429}
\left(\frac{\beta}{0.9}\right)^{0.771}
M_{1}^{-0.242}. 
\end{eqnarray}
Hence, we can see that although the pressure term is not purely expressible
as a function of $q$ alone, the various parameters are either expected to
remain reasonably constant between systems (eg. $\mu_{\rm H}$) or enter in  
with a weak dependency such as  $M_{1}^{-0.242}$.
As mentioned above, the observed distribution for {\it all} CVs is
$M_{1}=0.76\pm0.22$ which would produce a variation of only $\sim7\%$ in the
predicted pressure contribution. The dwarf nova subsample had an even smaller
range for $M_{1}$. 

\begin{figure}
\begin{center}
\includegraphics [angle=270,scale=0.5] {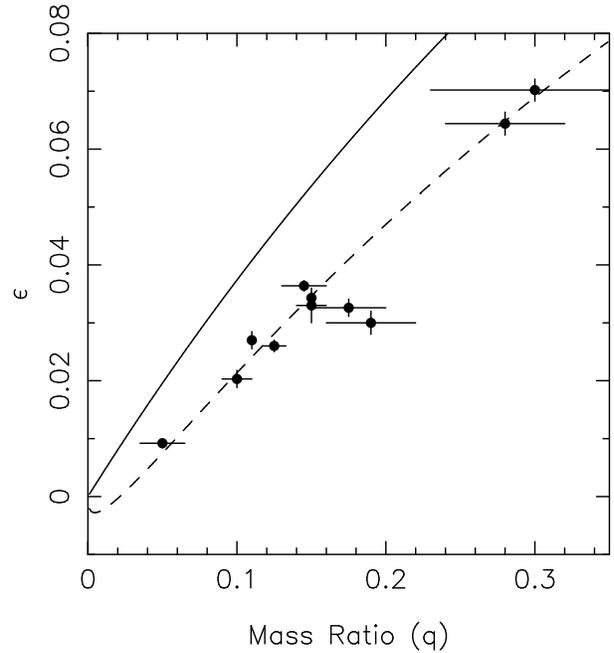}
\caption{Comparison of the calibration systems from 
\protect\citet{patterson05} with a model with purely dynamical precession 
(solid) and including the detail pressure term in 
equation~(\ref{eqn:finompress}) (dashed).}
\protect\label{fig:finevq}
\end{center}
\end{figure}

The final expression for the precession rate was fitted to the calibration 
systems of \citet{patterson05} using a single free parameter as the
constant of proportionality to the
functional form of $q$ in (\ref{eqn:finompress}). The best fitting value of 
$\eta_{\rm B}=0.0109$ gives $\frac{\chi^{2}}{N_{\rm obs}-1}=1.04$. With the 
typical values used 
above this implies $i=61^{\rm \circ}$. A comparison with the data is
plotted in Figure~\ref{fig:finevq}.  For the lowest $q$ systems 
the effect of the 
pressure term actually makes $\epsilon$ negative, ie. force the precession
to become retrograde. 

Inverting the process for 
systems of known $\epsilon$ but unknown $q$ allows us to derive values for 
the 88 systems in Table~9 of \citet{patterson05}. These are listed in 
Table~\ref{tab:qded}. The errors quoted in the table reflect the observational
errors propogated as appropriate but not the systematic errors that may arise
from variations of the parameters in the \citet{smith98} relations.
It is apparent that some systems have predicted values of $q$ in excess
of the expected limit of $q_{\rm max}=0.28$ for $\beta=0.9$. The largest 
derived value of
$q=0.437$ can be accommodated if $\beta=0.94$. If the resonance were to 
transition to 4:3 at some $q_{\rm max}$, the derived values of $q$ would be
even higher. However, axisymmetric structure models were used in
the above derivation. As $r_{j}$ approaches $R_{\rm T}$ this will 
become an increasingly invalid assumption. It might be reasonably expected then
that these high $q$ systems have true mass ratios close to $q_{\rm max}$
although, as we have seen, this limit is not well determined. 

It is also worth noting just 
how close is the derived $M_{1}$ in EG~Cnc to the Chandrasekhar limit. 
Although the mass-period relation we have used (equation~\ref{eqn:massradrel})
was derived from observations, there are strong theoretical grounds for 
expecting a significantly different relation for the lowest mass secondaries
once they become degenerate or semi-degenerate. EG~Cnc is towards the low-end 
in the range of $M_{2}$ where this effect may be important and we may be
extrapolating the relation beyond its range of validity. A 
non-main-sequence star would be expected to have a larger radius for the same 
mass as a main-sequence equivalent. This would lead to a larger pressure
effect for a given value of $q$. As a result we would tend to underestimate
both $M_{2}$ and $q$ when such deviation becomes important. In principle,
analytic forms for the mass-radius relations for the appropriate ranges of
$M_{2}$ could be included in place of (\ref{eqn:massradrel}).

\begin{figure}
\begin{center}
\includegraphics [angle=270,scale=0.5] {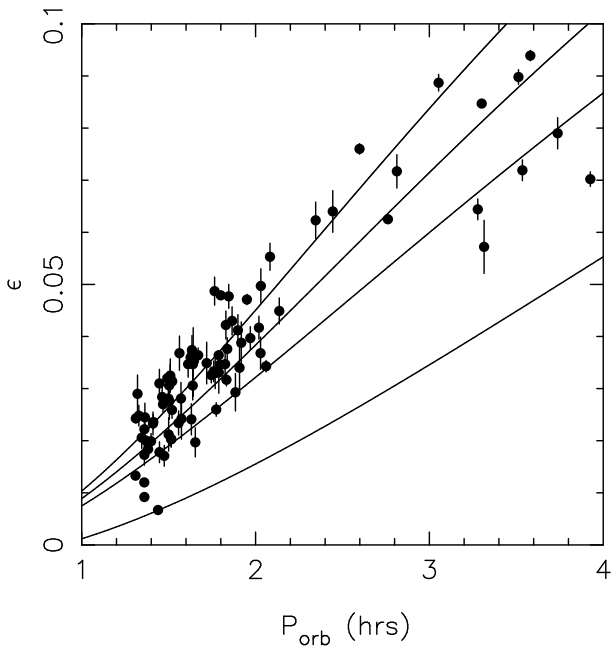}
\caption{Comparison of the observed superhump data from 
\protect\citet{patterson05} with a model  including the 
full pressure effect. The curves use 
the mean ($0.69$) and $\pm1\sigma$ ($0.01$) values for 
$M_{1}$ found for {\it dwarf novae} and in the 
$M_{2}$--$P_{\rm orb}$ relation  
from \protect\citet{smith98} as well as a further line with $M_{1}=1.44$. 
$\epsilon$ decreases with increasing $M_{1}$.
}
\protect\label{fig:epspcompfull}
\end{center}
\end{figure}

\begin{table*}
\protect\label{tab:qded}
\begin{center}
\begin{tabular}{lccccc} \hline
 Name & $P_{\rm orb}$(d) & $\epsilon$ & $q$ & $M_{2}$ & $M_{1}$ \\ \hline
    DI UMa      &  0.05456( 1) &   0.0133( 6) &   0.071( 2) & 0.058 & 0.821(25) \\
    V844Her     &  0.05464( 1) &   0.0243( 9) &   0.111( 3) & 0.058 & 0.526(16) \\
    LLAND       &  0.05505( 1) &   0.0290(36) &   0.128(14) & 0.059 & 0.459(49) \\
    SDSS0137-09 &  0.05537( 4) &   0.0248(20) &   0.113( 7) & 0.060 & 0.528(35) \\
    ASAS0025+12 &  0.05605( 5) &   0.0206(21) &   0.097( 8) & 0.061 & 0.624(49) \\
    AL Com      &  0.05667( 3) &   0.0120( 7) &   0.066( 3) & 0.062 & 0.933(35) \\
    WZ Sge      &  0.05669( 1) &   0.0092( 7) &   0.056( 3) & 0.062 & 1.100(49) \\
    RX1839+26   &  0.05669( 5) &   0.0173(20) &   0.085( 7) & 0.062 & 0.724(61) \\
    PU CMa      &  0.05669( 5) &   0.0222(20) &   0.103( 7) & 0.062 & 0.599(43) \\
    SW UMa      &  0.05681(14) &   0.0245(27) &   0.112(10) & 0.062 & 0.556(50) \\
    HV Vir      &  0.05707( 1) &   0.0200( 9) &   0.095( 3) & 0.062 & 0.657(23) \\
    MM Hya      &  0.05759( 1) &   0.0184(10) &   0.089( 4) & 0.063 & 0.710(29) \\
    WX Cet      &  0.05829( 4) &   0.0199(15) &   0.095( 5) & 0.065 & 0.682(39) \\
    KV Dra      &  0.05876( 7) &   0.0233(22) &   0.107( 8) & 0.065 & 0.610(46) \\
    T Leo       &  0.05882( 1) &   0.0236(14) &   0.108( 5) & 0.066 & 0.605(29) \\
    EG Cnc      &  0.05997( 9) &   0.0067( 8) &   0.047( 3) & 0.068 & 1.433(88) \\
    V1040 Cen   &  0.06028(10) &   0.0310(27) &   0.136(10) & 0.068 & 0.501(38) \\
    RX Vol      &  0.06030(20) &   0.0178(20) &   0.087( 7) & 0.068 & 0.782(65) \\
    AQ Eri      &  0.06094( 6) &   0.0284(21) &   0.126( 8) & 0.069 & 0.549(34) \\
    XZ Eri      &  0.06116( 1) &   0.0270(16) &   0.121( 6) & 0.070 & 0.576(28) \\
    CP Pup      &  0.06145( 6) &   0.0171(20) &   0.085( 7) & 0.070 & 0.830(71) \\
    V1159 Ori   &  0.06218( 1) &   0.0320(11) &   0.140( 4) & 0.072 & 0.512(15) \\
    V2051 Oph   &  0.06243( 1) &   0.0281(25) &   0.125( 9) & 0.072 & 0.576(43) \\
    V436 Cen    &  0.06250(20) &   0.0212(32) &   0.099(12) & 0.072 & 0.725(85) \\
    BC UMa      &  0.06261( 1) &   0.0306(14) &   0.134( 5) & 0.072 & 0.538(21) \\
    HO Del      &  0.06266(16) &   0.0276(35) &   0.123(13) & 0.072 & 0.588(63) \\
    EK TrA      &  0.06288( 5) &   0.0321(25) &   0.140(10) & 0.073 & 0.519(35) \\
    TV Crv      &  0.06290(20) &   0.0325(32) &   0.142(12) & 0.073 & 0.513(44) \\
    VY Aqr      &  0.06309( 4) &   0.0203(15) &   0.096( 5) & 0.073 & 0.761(43) \\
    OY Car      &  0.06312( 1) &   0.0203(15) &   0.096( 5) & 0.073 & 0.761(43) \\
    RX1131+43   &  0.06331( 8) &   0.0259(16) &   0.117( 6) & 0.074 & 0.630(32) \\
    ER UMa      &  0.06336( 3) &   0.0314(11) &   0.138( 4) & 0.074 & 0.536(17) \\
   DM Lyr       &  0.06546( 6) &   0.0281(31) &   0.125(12) & 0.078 & 0.620(58) \\
    UV Per      &  0.06489(11) &   0.0234(23) &   0.108( 8) & 0.077 & 0.711(56) \\
    AK Cnc      &  0.06510(20) &   0.0368(33) &   0.158(13) & 0.077 & 0.486(40) \\
    AO Oct      &  0.06557(13) &   0.0242(39) &   0.111(14) & 0.078 & 0.704(92) \\
    SX LMi      &  0.06717(11) &   0.0347(25) &   0.150(10) & 0.081 & 0.538(35) \\
    SS UMi      &  0.06778( 4) &   0.0360(15) &   0.155( 6) & 0.082 & 0.528(20) \\
    KS UMa      &  0.06796(10) &   0.0241(30) &   0.110(11) & 0.082 & 0.747(75) \\
    V1208 Tau   &  0.06810(20) &   0.0374(28) &   0.161(11) & 0.083 & 0.514(35) \\
    RZ Sge      &  0.06828( 2) &   0.0306(28) &   0.134(11) & 0.083 & 0.617(49) \\
    TY Psc      &  0.06833( 5) &   0.0347(15) &   0.150( 6) & 0.083 & 0.553(21) \\
    IR Gem      &  0.06840(30) &   0.0351(66) &   0.152(26) & 0.083 & 0.548(93) \\
    V699 Oph    &  0.06890(20) &   0.0197(28) &   0.094(10) & 0.084 & 0.895(97) \\
    CY UMa      &  0.06957( 4) &   0.0364(14) &   0.157( 5) & 0.085 & 0.545(19) \\
    FO And      &  0.07161(18) &   0.0349(40) &   0.151(16) & 0.089 & 0.592(61) \\
    OU Vir      &  0.07271( 1) &   0.0326(15) &   0.142( 6) & 0.092 & 0.643(26) \\
    VZ Pyx      &  0.07332( 3) &   0.0333(20) &   0.145( 8) & 0.093 & 0.641(34) \\
    CC Cnc      &  0.07352( 5) &   0.0487(27) &   0.207(11) & 0.093 & 0.449(25) \\
    HT Cas      &  0.07365( 1) &   0.0330(30) &   0.144(12) & 0.093 & 0.651(52) \\
    IY UMa      &  0.07391( 1) &   0.0260(13) &   0.117( 5) & 0.094 & 0.802(33) \\
    VW Hyi      &  0.07427( 1) &   0.0331( 8) &   0.144( 3) & 0.095 & 0.658(13) \\
    Z Cha       &  0.07450( 1) &   0.0364( 9) &   0.157( 4) & 0.095 & 0.607(14) \\
    QW Ser      &  0.07453(10) &   0.0331(40) &   0.144(15) & 0.095 & 0.661(71) \\
    WX Hyi      &  0.07481( 1) &   0.0346(14) &   0.150( 5) & 0.096 & 0.639(23) \\
    BK Lyn      &  0.07498( 5) &   0.0479( 7) &   0.204( 3) & 0.096 & 0.471( 7) \\
   RZ Leo       &  0.07604( 1) &   0.0347(25) &   0.150(10) & 0.098 & 0.654(42) \\
    AW Gem      &  0.07621(10) &   0.0422(27) &   0.180(11) & 0.099 & 0.548(33) \\
    SU UMa      &  0.07635( 5) &   0.0317(12) &   0.139( 5) & 0.099 & 0.713(23) \\
\end{tabular}
\caption{Derived parameters for all the systems with measured period excesses
from \protect\citet{patterson05}. The errors on $q$ reflect the
observational errors propogated appropriately. $M_{2}$ has been calculated using 
equation
(\ref{eqn:m2prel}) and $M_{1}=\frac{q}{M_{2}}$. No allowance has been made for the 
systematic error that would arise from the errors in the parameters in this 
relation. Similarly, no error is quoted for $M_{2}$ since this is dominated by the assumption
of main sequence structure.}
\end{center}
\end{table*}

\addtocounter{table}{-1}
\begin{table*}
\begin{center}
\begin{tabular}{lccccc} \hline
 Star & $P_{\rm orb}$(d) & $\epsilon$ & $q$ & $M_{2}$ & $M_{1}$ \\ \hline
   SDSS1730+62  &  0.07655( 9) &   0.0376(22) &   0.162( 9) & 0.099 & 0.615(33) \\
    HS Vir      &  0.07690(20) &   0.0477(23) &   0.203(10) & 0.100 & 0.493(23) \\
    V503 Cyg    &  0.07770(20) &   0.0430(27) &   0.183(11) & 0.102 & 0.555(34) \\
    V359 Cen    &  0.07990(30) &   0.0388(40) &   0.166(16) & 0.106 & 0.639(61) \\
    CU Vel      &  0.07850(20) &   0.0293(36) &   0.130(14) & 0.103 & 0.798(83) \\
   NSV 9923     &  0.07910(20) &   0.0412(30) &   0.176(12) & 0.105 & 0.595(41) \\
    BR Lup      &  0.07950(20) &   0.0340(40) &   0.147(15) & 0.105 & 0.716(75) \\
    V1974 Cyg   &  0.08126( 1) &   0.0471(10) &   0.201( 4) & 0.109 & 0.544(11) \\
    TU Crt      &  0.08209( 9) &   0.0397(22) &   0.170( 9) & 0.111 & 0.653(34) \\
    TY PsA      &  0.08414(18) &   0.0417(22) &   0.178( 9) & 0.115 & 0.648(32) \\
    KK Tel      &  0.08453(21) &   0.0368(31) &   0.158(12) & 0.116 & 0.734(56) \\
    V452 Cas    &  0.08460(20) &   0.0497(33) &   0.212(14) & 0.116 & 0.550(37) \\
    DV Uma      &  0.08585( 1) &   0.0343(11) &   0.149( 4) & 0.119 & 0.801(23) \\
   YZ Cnc       &  0.08680(20) &   0.0553(26) &   0.236(12) & 0.121 & 0.514(25) \\
    GX Cas      &  0.08902(16) &   0.0449(25) &   0.191(11) & 0.126 & 0.660(37) \\
    NY Ser      &  0.09775(19) &   0.0623(35) &   0.268(16) & 0.146 & 0.546(33) \\
    V348 Pup    &  0.10184( 1) &   0.0640(40) &   0.276(19) & 0.156 & 0.565(39) \\
    V795 Her    &  0.10826( 1) &   0.0760(10) &   0.336( 5) & 0.172 & 0.512( 8) \\
    V592 Cas    &  0.11506( 1) &   0.0625( 5) &   0.269( 2) & 0.189 & 0.703( 6) \\
    TU Men      &  0.11720(20) &   0.0717(32) &   0.314(16) & 0.195 & 0.621(32) \\
    AH Men      &  0.12721( 6) &   0.0887(16) &   0.406( 9) & 0.222 & 0.546(13) \\
   DW Uma       &  0.13661( 1) &   0.0644(20) &   0.278(10) & 0.248 & 0.893(31) \\
    TT Ari      &  0.13755( 1) &   0.0847( 7) &   0.383( 4) & 0.251 & 0.655( 7) \\
    V603 Aql    &  0.13810(20) &   0.0572(51) &   0.244(23) & 0.252 & 1.032(97) \\
    PX And      &  0.14635( 1) &   0.0898(14) &   0.412( 8) & 0.277 & 0.671(13) \\
    V533 Her    &  0.14730(20) &   0.0719(20) &   0.315(10) & 0.279 & 0.888(29) \\
    BB Dor      &  0.14920(10) &   0.0939(10) &   0.437( 6) & 0.285 & 0.652( 9) \\
    BH Lyn      &  0.15575( 1) &   0.0790(30) &   0.352(16) & 0.305 & 0.868(40) \\
    UU Aqr      &  0.16358( 1) &   0.0702(14) &   0.306( 7) & 0.330 & 1.077(25) \\
\end{tabular}
\caption{{\bf Cont.}
Derived parameters for all the systems with measured period excesses
from \protect\citet{patterson05}.}
\end{center}
\end{table*}

As a disc outburst proceeds, we would expect $\dot{M}$ to steadily 
decrease. Since $c^{2}\propto\dot{M}^{\frac{3}{10}}$ this would cause the 
pressure term to also shrink and thus the period excess to increase
during an outburst.
Observationally, the opposite appears to be the case, with the  period excess
decreasing with time \citep{patterson93}. The only available parameter to
counter this is $\tan i$ implying that the pitch angle $i$ increases during an 
outburst. 

Although the treatment of \citet{lubow91} includes the coupling of the tides,
spiral waves and eccentricity, it shares the shortcoming of a purely dynamical 
resonance in that the properties are characterised by those at a single
radius. A recent preprint of a paper by \citet{goodchild06} has 
addressed this problem
with a detailed analysis solving the equations for the disc behaviour 
integrated over the whole range of disc radii. Their final equation describing
the generation, damping and dynamics of eccentricity in the disc is
\begin{eqnarray}
2r\Omega \frac{\partial E}{\partial t} & =
\frac{i E}{\rho} \frac{\partial p}{\partial r} +
\frac{i}{r^{2} \rho} \frac{\partial}{\partial r}
\left( \left(\gamma-i\alpha_{\rm b}\right) p r^{3} 
\frac{\partial E}{\partial r}\right)
+\nonumber & \\ 
& \frac{i q \Omega^{2} r^{3}}{2 d^{2}}
\left( b^{1}_{\frac{3}{2}}\left(\frac{r}{d}\right) E\right)
+2\xi r \Omega E \delta(r - r_{\rm res}). & \label{eqn:goodeqn}
\end{eqnarray}
Here $E$ is the complex eccentricity $E=e {\rm e}^{i \omega}$, $p$ 
and $\rho$ are the local disc pressure and density, $\gamma-i\alpha_{\rm b}$
is a complex adiabatic exponent, $\xi=2.08\omega_{\rm orb}q^{2} r_{\rm res}$ 
is the eccentricity growth rate for a resonance at radius $r_{\rm res}$ from
\citet{lubow91} and $\Omega$ is the angular velocity of the orbiting disc
material. Unsuprisingly, the full solution has to be found numerically.
These authors do so with undisturbed (vertically integrated) structure 
distributions given by
\begin{eqnarray}
P & = & P_{\rm sc} \left(\frac{r}{r_{\rm sc}}\right)^{-\frac{3}{2}}
\left(1\sqrt\frac{r_{\rm in}}{r}\right) 
\tanh\left(\frac{r_{\rm out}-r}{\nu r_{\rm sc}}\right) \label{eqn:PGood}\\
\Sigma & = & \Sigma_{\rm sc} \left(\frac{r}{r_{\rm sc}}\right)^{-\frac{3}{4}}
\left(1\sqrt\frac{r_{\rm in}}{r}\right)^{0.7}
\tanh\left(\frac{r_{\rm out}-r}{\nu r_{\rm sc}}\right) \label{eqn:SigmaGood}
\end{eqnarray}
where $r_{\rm out}$ and $r_{\rm in}$ are the outer and inner radii disc radii
and $P_{\rm sc}$ and $\Sigma_{\rm sc}$ are the pressure and surface density
at the scaling radius $r_{\rm sc}$. Considering the 
first bracketed terms in either case, the \citet{cannizzo92a} equations
used above have the same radial dependence. The second bracketed 
terms reflect the behaviour close to the inner radius. The final terms
were chosen to implement the boundary conditions at the outer edge of the
disc. Since we would expect the inner and outer terms to only become important
close to the limiting mass ratios and given the similarity otherwise of the 
radial dependences of the equations to those used earlier, we do not
need to repeat the integration here but compare the final results 
produced.

The results in Figure~8 of \citet{goodchild06} show a similar excursion to 
a negative period excess for small $q$. The curves also turn up to high
$\epsilon$ as they assymptotically approach $q_{\rm max}$ which our 
expression does  not produce. However, the 
turn off occurs very close to $q_{\rm max}$ for the best fitting curve
and the details of the behaviour here depend on the functional form chosen
to implement the boundary conditions for the outer edge of the disc in
equations (\ref{eqn:PGood}) and (\ref{eqn:SigmaGood}). Away from the extreme
mass ratios the two sets of results are very close.

\subsection{Empirical Fit}

\begin{figure}
\begin{center}
\includegraphics [angle=270,scale=0.5] {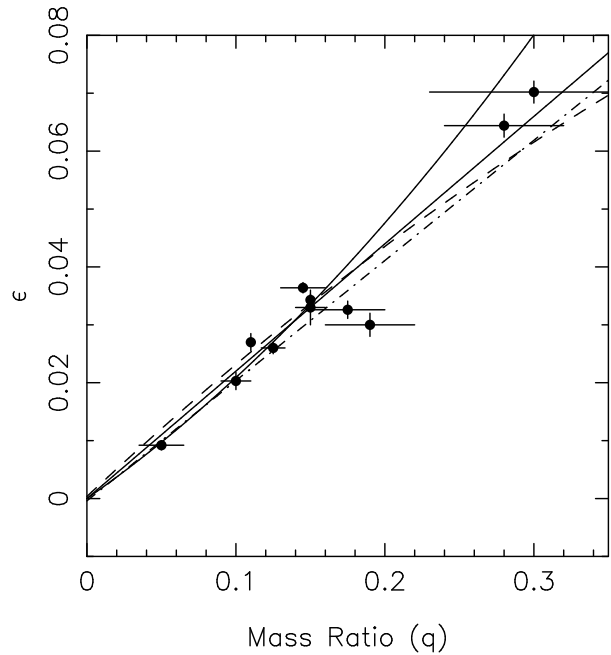}
\caption{Comparison of the observed superhump data from 
\protect\citet{patterson05} with the polynomial approximations 
for empirical fits  (equations \protect\ref{eqn:pattlinapp} and 
\protect\ref{eqn:pattquadapp}) (solid),
 this work (\ref{eqn:ourquad}) (dashed) and full integration 
(\ref{eqn:goodlin}) (dot-dashed).
}
\protect\label{fig:empcomp}
\end{center}
\end{figure}

In an attempt to calibrate an empirical relation between $\epsilon$ and $q$,
\citet{patterson05} extended the earlier fit of
\begin{equation}
\epsilon=0.22q \label{eqn:pattlinapp}
\end{equation}
\citep{patterson01} to
\begin{equation}
\epsilon=0.18q +0.29q^{2}. \label{eqn:pattquadapp}
\end{equation}
We can view these as phenomenologically derived equivalents to the Maclaurin 
series for our analytic expression. Given the complexity of the 
full expressions, 
however, rather than carry out the necessary differentiation, it is simplest 
to generate $\epsilon(q)$ predictions from our formula numerically
and then find the best fitting polynomial to these values using standard
methods \citep{numrec}. This approach gives 
an approximate formula
\begin{equation}
\epsilon=3.5\times10^{-4}+0.24q-0.12q^{2} \label{eqn:ourquad}
\end{equation}
over the range $0.01<q<0.4$. This expression differs both with a very small
constant offset and with a quadratic term that 
causes a curvature in the opposite sense to the empirical expression. This
can be attributed to removing the $q_{\rm max}$ limitation imposed by
U~Gem for  BB~Dor. \citet{goodchild06} derive a best fit formula for their
numerical integrations of (\ref{eqn:goodeqn}) of
\begin{equation}
\epsilon=-4.1\times10^{-4}+0.2076q. \label{eqn:goodlin}
\end{equation}
These polynomial forms are compared graphically in Figure~\ref{fig:empcomp}.
The linear empirical fit, the full integration polynomial and our result
all appear in good agreement.

\section{Summary}
We have shown that the standard dynamical method of calculating the precession 
rate of superhumping CVs with a 3:2 provides such a poor fit to the data that 
a 4:3 resonance is actually a better fit!
We have confirmed the importance of including the pressure related term in
the calculation of the precession rate \citep{murray00} 
which fits the data better than any pure resonance model. The pressure term 
has been reduced to a function of $q$ and the total, analytic precession rate
shown to be equivalent to the
empirically derived expressions of \citet{patterson05}. These anayltic
expressions also produce precession rates in good agreement with those
from the detailed integrations carried out by \citet{goodchild06}.
This formulation has
been used to calculate values for $q$ in systems which would otherwise 
be unknown.

\section*{ACKNOWLEDGEMENTS}

I thank Robert Hynes and Juhan Frank for helpful remarks and advice regarding 
the work in this paper and the anonymous referee for insightful
remarks that improved the presentation of these results.
\\

\end{document}